\documentclass{article}
\usepackage{dcase2018,amsmath,graphicx,url,times,booktabs, tabularx}
\usepackage{multirow,multicol}

\title{An extensible cluster-graph taxonomy for\\ open set sound scene analysis}

\name{Helen L. Bear and Emmanouil Benetos\thanks{This work was funded under EPSRC grant EP/R01891X/1. EB is supported by a RAEng Research Fellowship (RF/128).}}

\address{School of Electronic Engineering and Computer Science, Queen Mary University of London, UK\\
\texttt{\{h.bear, emmanouil.benetos\}@qmul.ac.uk}}

\begin{document}
\ninept
\maketitle
\begin{sloppy}
\begin{abstract}
We present a new extensible and divisible taxonomy for open set sound scene analysis. This new model allows complex scene analysis with tangible descriptors and perception labels. Its novel structure is a cluster graph such that each cluster (or subset) can stand alone for targeted analyses such as office sound event detection, whilst maintaining integrity over the whole graph (superset) of labels. The key design benefit is its extensibility as new labels are needed during new data capture. Furthermore, datasets which use the same taxonomy are easily augmented, saving future data collection effort. We balance the details needed for complex scene analysis with avoiding `the taxonomy of everything' with our framework to ensure no duplicity in the superset of labels and demonstrate this with DCASE challenge classifications. 
\end{abstract}

\section{Introduction}
\vspace{-0.5em}
\label{sec:intro}
In sound scene analysis, that is describing a scene and its constituent events from an audio input, most work poses the problem as a closed set problem \cite{aly2005survey}. This means researchers use a defined set of class labels with various levels of confidence. In doing so, many datasets and associated taxonomies/ontologies have been created \cite{salamon2014dataset,gemmeke2017audio,giannoulis2013detection}. These approaches are based on the assumption that sound scenes can be described from a finite collection of labels. However, particularly for complex real-world scenes, the problem is more akin to an open set classification task \cite{battaglino2016open,macqueen1967some}. That is, there are infinite possible descriptor labels, and furthermore, many combinations of labels possible. 
The range of tasks in computational sound scene analysis is varied (e.g. scene classification, event recognition) and each time a new dataset is created, we also create new sets of labels. This reduces the reusability and value of data which is expensive and time-consuming to collect and annotate. 
Therefore, we seek a new structure taxonomy to support the research community, and thus we need a class labeling mechanism which can:
\begin{itemize}
\item extend as the complexity of scenes develops and research spreads into new scenes of interest,
\item not duplicate descriptors across sub-areas of a taxonomy,
\item be divided up for tackling nuanced sub-problems in sound scene analysis, and
\item enable multi-perspective descriptors; that is, how a human perceives a scene rather than physical logical descriptors.
\end{itemize}
This paper provides a framework for a modifiable and extensible sound scene taxonomy for \emph{all} scenes, where  one can analyse a set of any events in any environment. The result is scenes that are describable by a set of descriptors consistent across datasets. Descriptors are singular scene labels which can describe: the environment, events, or the context (how a human could perceive the scene).

The rest of this paper is structured as follows: we discuss the background of taxonomy development, before presenting our extensible taxonomy architecture and a framework for populating it with labels and maintaining it. We then demonstrate it using the label sets provided with the DCASE challenge label sets from 2013--2018 before summarising the benefits of this new approach. 

\section{Background}
\vspace{-0.5em}
\label{sec:back}
We begin with the following definitions: 
\\ \indent a \textbf{taxonomy} is \textit{a scheme of classification},
\\ \indent a \textbf{label set} is \textit{a collection of class names},
\\ \indent an \textbf{ontology} is \textit{a set of concepts and categories in a domain that shows their properties and the relations between them} and,
\\ \indent a \textbf{thesaurus} is \textit{a book that lists words in groups of synonyms and related concepts}.

Our proposal is an extensible taxonomy which combines an ontology to organise label sets supported by a thesaurus to avoid duplication or misnomers between related subsets of labels. In doing so, we organise a collection of label sets into a graph structure to enable relations between class labels and subsets of labels. This moves us away from using hierarchies to organise classification labels. Benefits of doing so allows a user to have specificity and precision at multiple levels for both scenes and events, and researchers can simply share lists of edges formed by label pairs to share the whole taxonomy or part required thereof.  

To do this, we build upon prior taxonomy work in sound scene analysis. We first discuss soundscapes, then events as these are typically (not always) addressed independently. Our third section reviews joint attempts. As we discuss this prior work, we recall that the purpose of sound scene analysis can vary. The requirements of a taxonomy which is fit for urban events, are unlikely to be suitable for urban scenes without some processing or modification. Therefore, as we discuss previous works we focus on their primary goal, before addressing modification requirements to aid an extensible taxonomy for complex scenes.

\textit{Soundnet} \cite{aytar2016soundnet} was developed using transfer learning from computer vision research. Scene understanding is a major computer vision research topic including for example object recognition and semantic understanding. Using this prior knowledge, transfer learning enabled deduction of acoustic labels. \textit{AudioSet} \cite{gemmeke2017audio} is a two tier hierarchical ontology of $632$ audio classes based on prior literature and using youtube video clips. \textit{Urban sounds} \cite{salamon2014dataset} groups with four top level groups: human, nature, mechanical and music. Here the authors target urban soundscapes and determine that leaf labels must be specific, e.g. car ``brakes'', ``engine'' or ``horn'', instead of simply ``car''. However, this means that the leaf labels are specific for the clusters and cannot be shared or compounded with others for complex scenes outside the remit of urban environments.

Gaver's taxonomy \cite{gaver1993world} was evaluated by Houix \cite{houix2012lexical} and found that sounds can be cross-classified by alternative principles depending on sound presentation and the background of a listener. This is supported by Guyot \cite{guyot1997chapitre} who observed a distinction in the classification labels used between acoustician/non-acoustician listeners. In further human classification experiments, Van der Veer \cite{vanderveer1980ecological} and Shubert \cite{schubert1975role} demonstrated that when classifying sounds, humans tend to use compounds of objects and actions. This is evidence that a wholly encompassing open set taxonomy which permits compound descriptors from fundamental labels would be beneficial for fair audio assessment. One approach in \cite{heittola2013context} uses a hierarchy between the environment and the scene. The classification strategy uses context (environment by our definition) to aid event detection similar to human comprehension methods. Within their own dataset, environments are distinct, e.g. beach, park, on a bus. 
In addition, Pijanowski \textit{et al.} \cite{pijanowski2011soundscape} used three abstract groupings; geophony, biophony, and anthophony for environmental/outside spaces. Similarly, Delage in \cite{amphoux1997paysage} used three abstract groups for urban sounds: sounds of nature, direct human activity, and indirect human activity. These are examples of clusters/groups of labels within a whole set and we use this model as inspiration for our cluster-graph ontology (as we present in Sec.~\ref{sec:graphTax}). 

Of these taxonomies, many use abstract labels for grouping types of labels. Yet, the labels in these groups can duplicate across groupings which creates confusion across datasets and research problems. Homographs \textit{(def: each of two or more words having the same spelling or pronunciation but different meanings and origins)} and synonyms \textit{(def: a word or phrase that means exactly or nearly the same as another word or phrase in the same language)} \cite{simpson1989oxford} are troublemakers in developing taxonomies for complex sound scenes. Compounded with the `extra' abstract labels, we witness inaccuracies and ambiguous classifications. Therefore, we need a set of rules for extending the taxonomy for each cluster in the graph. 
Some prior work does not restrict itself to only Urban/Rural sounds or Scene/Event sound schemes. A number of schemes have wide reaching coverage, namely \cite{guyot1997chapitre,heittola2013context,gaver1993world,houix2012lexical}. Importantly, not all taxonomies fit into this simple cross-referencing structure, thus any future solution must enable cross problem label sharing for complex scenes. 
Also, any new taxonomy should address the human ability to `cross-classify', i.e. enabling a machine to classify scenes and their components in multiple variants. That is, one class might exist in multiple different scenes, and the best approach for recognising that class can vary by the scene or context. 

\textbf{The first thesaurus: Wordnet} \cite{wordnet} is an open-source online thesaurus containing groups of synonyms named \textit{synsets}. Synsets are indexed by the preferred term, that is the word most commonly used. Using this we can ensure that our cluster-graph has one single label per leaf and that highest usage synonyms are used to reduce duplicity and to manage homographs. All punctuation should be removed, and labels saved in lower case. 

\section{A graph taxonomy}
\vspace{-0.5em}
\label{sec:graphTax}
In Sec.~\ref{sec:back} we stated our proposal is an extensible taxonomy which combines a graph-based ontology to organise label sets supported by a thesaurus to avoid duplication or misnomers between related subsets of labels. This means that the structure and relations between class labels are modelled as a graph, where class labels are grouped in clusters of leaf nodes, label sets are subgraphs, and the relations between these are edges. As we build upon label sets already provided, this will be a cluster graph \cite{west2001introduction}. This structure is new for sound scene analysis taxonomy, although graphs have been used in vision tasks \cite{sridhar2010discovering}. 

The graph taxonomy enables one to make cuts through the total set of all labels/leaf nodes such that it can be sliced into relevant subsets of labels for different challenges in sound scene analysis. Each cluster is related via the graph schema so enables use of the same categories presented in prior datasets (reorganised) into an cluster graph. This also encourages data augmentation (e.g. \cite{bai2017automatic}) across clusters so that researchers can quickly amalgamate bigger datasets without the collection and annotation costs. Depending on raw data used, intra-cluster distance measures can be calculated for multi-label classification methods.

A significant benefit of this architecture is that it can be represented in both graph and set form, more commonly known as a Venn diagram \cite{enderton1977elements}. With both we can show both the detail of the graph and the paths between labels, but also the most common ways of cutting the graph (based on prior works) into subsets, or sub-graphs. 

In this taxonomy we use the following terms as defined: 
\indent \textbf{Environment}: tangible description of a scene, e.g. indoors, light, building, people.\\
\indent \textbf{Context}: a human perceptual description of a scene, e.g. meeting, party, sea side.\\
\indent \textbf{Event}: sound emitting actions or objects in the scene, e.g. speech, writing, keyboard presses, walking. 
\begin{figure}[!htb]
\centering
\includegraphics[width=.7\columnwidth,keepaspectratio]{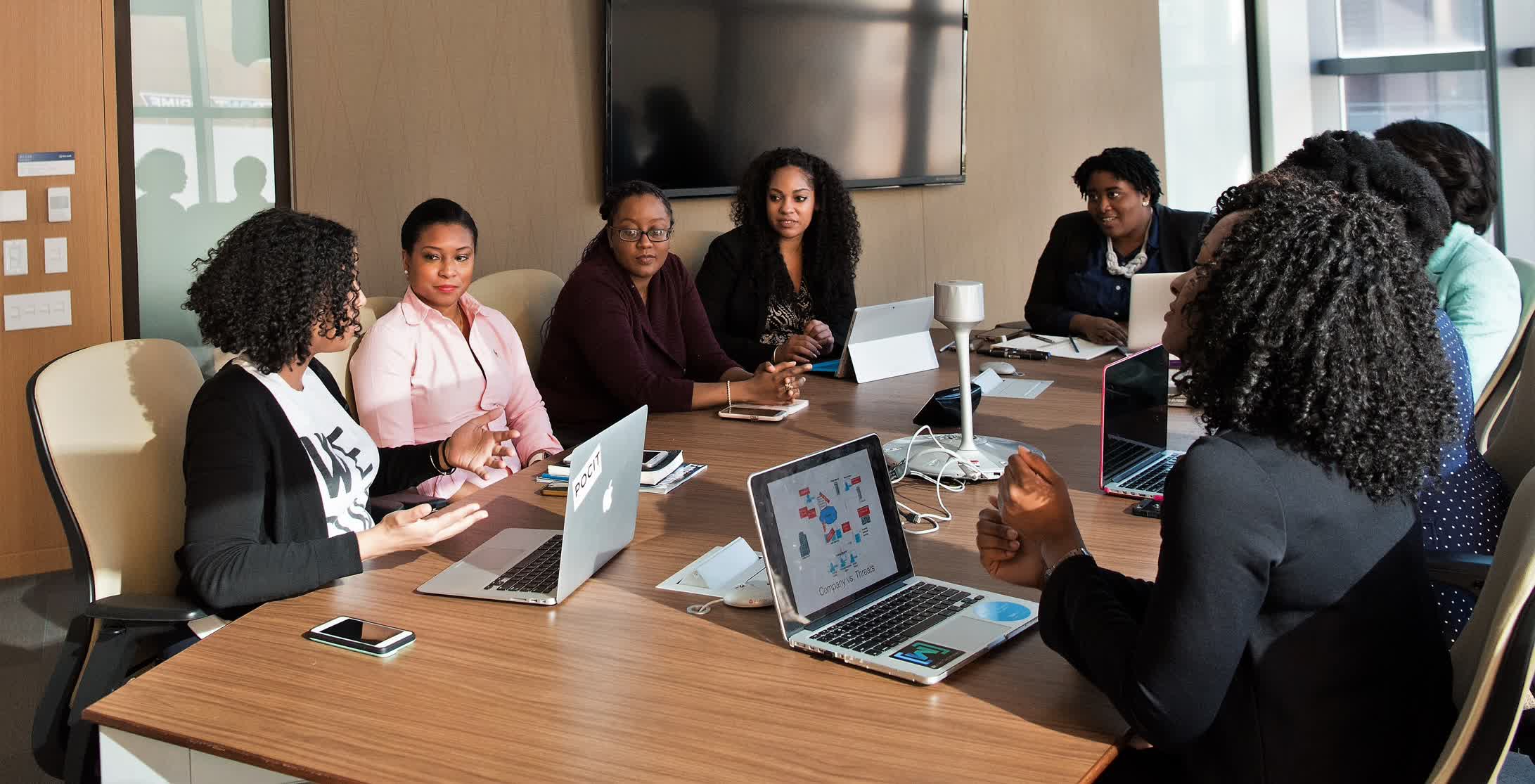} \\
\vspace{.5em}Example Scene \\ 
\vspace{.5em} \includegraphics[width=.7\columnwidth]{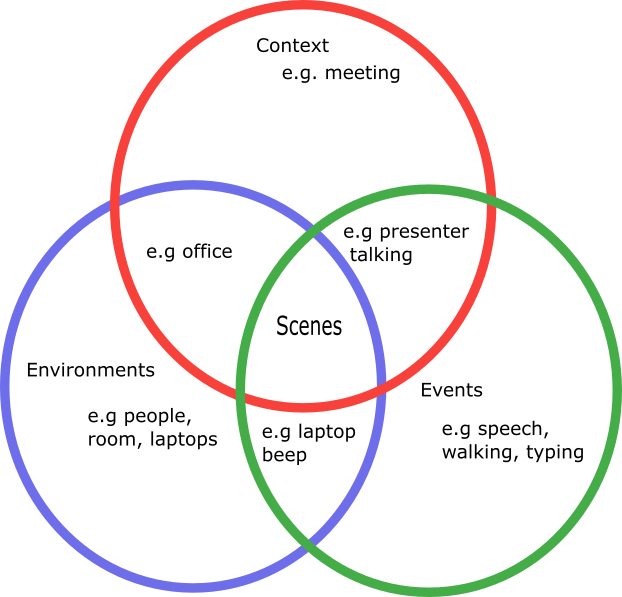}
\caption{Example scene (top) \cite{copyrightFree} and first graph cuts for subsets of label sets (bottom).}
\label{fig:labelHierachy}
\end{figure}

These terms form the fundamental ontology of label sets as shown in Fig.~\ref{fig:labelHierachy}. This Venn diagram shows boundaries between the label subsets (labels and the subgroupings are graph nodes), and how the subsets relate to each other to form alternative descriptors. The example labels are examples for a scene similar to Fig.~\ref{fig:labelHierachy}(top).

Thus we begin with a superset $T$, which is a set of all labels, whether part of the environment $en$, event $ev$, or context $c$ subsets. At this top level, all labels in $T$ can be related to all others which forms the full graph $G=(T,E_t)$ where $E_t$ is the set of edges between all pairs of $T$. With $T$, many subsets $t$ of $T$ can be created for any sound scene problem (urban, office events etc), with the reassurance that a consistent $T$ enables extension of current taxonomies. 

 
\subsection{The ontology $O$}
\vspace{-0.25em}
The remaining element of this taxonomy is forming an ontology of the relations between subsets. We observed in prior work both the use of compound labels and also the human tendency to describe sound scenes perceptually. The compound labels form what is seen in Figure~\ref{fig:labelHierachy} as overlap subsets between the event and environment subsets, and the human perception labels are the context subset. 

Collectively, these subsets are the abstract classes used in prior taxonomies but now we can also use them as subset labels to support multi-label analysis without needing data annotated at the label level in the leaf nodes, reducing the annotation overhead and maximising usability of the data already available. 

Subset names are \textbf{not} included as nodes (labels) in the graph. These are maintained as a separate list. It is essential that there is no duplication over all subsets, that is the same label can appear in more than one subset, but only as itself, not a synonym. It is also essential that $T$ contains all unique elements of all subsets. This constrains $T$ to prevent duplicity and unnecessary complexity.  

\subsection{Sub graph architecture}
\vspace{-0.25em}
Because our ontology architecture is a graph, we have the significant benefit that we can make many graph cuts through the total set of labels ($T$) to create new sub-graphs, or new label sets (as we will see in Sec.~\ref{sec:initialisation}). The obvious subgroups are the pre-populated clusters building on the label sets already available in literature. Therefore, we avoid using abstract names for labels. Rather, these should be a suitable name for an aggregate set, incorporating the super/sub class relation from a hierarchy of previous taxonomies without creating a node identifier that is not suitable as a label, e.g. `subway train' is a composite of `subway' and `train' which stand alone as labels.

Different complex, polyphonic sound scenes require variable numbers of categories to describe the scene. Using our Fig~\ref{fig:labelHierachy} example, some would relate to the environment e.g. `office', some events (or actions) like `talking', and some objects, `laptops'. Given the volume of possible combinations, we suggest using label vectors to represent a set of categories per scene or compound labels for measuring accuracy, rather than creating new ones. 


The use of context to improve event classification has been seen as beneficial in sound scene analysis \cite{heittola2013context}; for example, if we are confident the scene is a beach, we raise the probability of events being waves or walking on sand. An alternative example is, the lack of an adverse class infers a class to be true, meaning if it is not raining nor windy then the probability of a sunny scene is more likely. This is where the clusters of label sets need weighted relations between labels within a subset or, relations (again weighted) between sets of labels. Any weights would be subject to data selection. 

Each cluster can be treated as both its own ontology, that is each label in a cluster set will relate to the other labels in the cluster, as such, distances between the labels might be calculated. Furthermore, as the taxonomy extends, distances between clusters, or pairs of leaves in different clusters, can be used to improve the confidence of multi-label classification predictions.

\section{New Label Framework}
\vspace{-0.5em}
In order to add to our graph, we present a framework (Fig.~\ref{fig:additionProcess}) to maintain graph integrity. We use this with the DCASE challenge labels to initialise $T$. 
\vspace{-0.15in}
\begin{figure}[!h]
\centering
\includegraphics[width=\columnwidth,keepaspectratio]{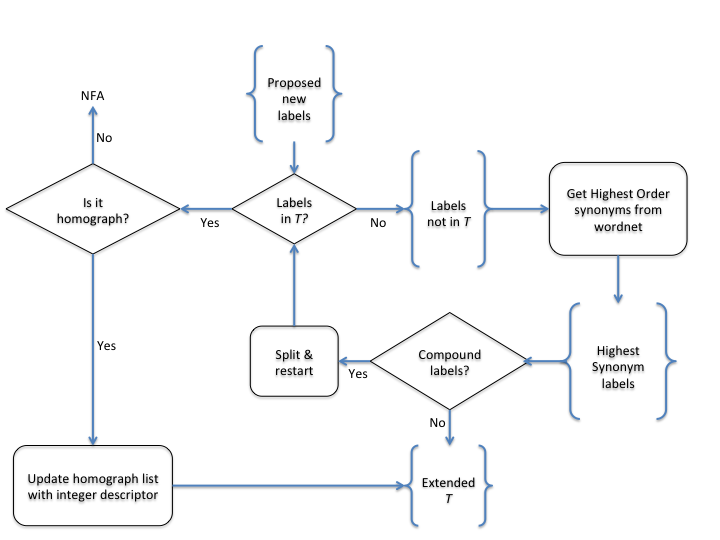}
\caption{Framework for extending the taxonomy.}
\label{fig:additionProcess}
\end{figure}

\section{Initialisation with DCASE examples}\label{sec:initialisation}
\vspace{-0.5em}
\begin{table*}[!ht]
\centering \caption{Event label set clusters from various DCASE challenges (2013,2016-2018)}
\begin{tabular}{lllllll}
DCASE 13' (T2\&3) 		& DCASE 16'(T2)	 	& DCASE 16'(T3) 		& DCASE 17'(T3) 	& DCASE 17'(T4) 		& DCASE 18' (T4) \\
\hline \hline
door knock, 			& clearing throat 	& (object) rustling		& brakes squeaking 	& train horn			& speech\\
door slam,				& coughing 			& (object) snapping		& car				& air horn, truck horn	& dog\\
speech, 				& door knock 		& cupboard 				& children			& car alarm				& cat\\
laughter, 				& door slam 		& cutlery				& large vehicle		& reversing beeps		& alarm \\
keyboard clicks, 		& drawer (close) 	& dishes				& people speaking	& ambulance (siren)		& dishes\\
objects hitting table, 	& human laughter 	& drawer				& people walking	& police car (siren)	& frying\\
keys clinging, 			& Keyboard 			& glass jingling 		&					& fire engine			& blender\\
phone ringing, 			& keys (put on table) & object impact		&					& civil defense siren 	& running water\\
turning page, 			& page turning 		& people walking		& 					& screaming				& vacuum cleaner\\
\cline{5-5}
cough, 					& phone ringing 	& washing dishes		&					& bicycle  				& electric shaver/toothbrush\\
printer, 				& speech 			& water tap running		&					& skateboard\\
\cline{3-3}
short alert-beeping, 	& 					& (object) Banging 		&					& car \\
clearing throat, 		& 					& bird singing			&					& car passing by\\
mouse click,			& 					& car passing by 		&					& bus \\
drawer 					& 					& children shouting		&					& truck \\
switch	 				& 					& people speaking 		&					& motorcycle \\
						&					& people walking 		&					& train\\
                      	& 					& wind blowing 			&					&\\  
\hline
\end{tabular}
\label{tab:dcaseTaxes_events}
\end{table*}

We use all the label sets from DCASE event challenges 2013 to 2018 \cite{giannoulis2013detection,mesaros2018detection,mesaros2017dcase,dcase2018} as listed chronologically from left to right in Table~\ref{tab:dcaseTaxes_events}, and our framework to demonstrate how these produced $T$ \footnote{$T$ and the first subsets: $en$, $ev$, $c$ are available from \texttt{soundscape.eecs.qmul.ac.uk/the-extensible-taxonomy/}}. We denote each set in the format `DxxTy' meaning D=DCASE, xx=the year, T=Task, and y=the task number.
Working from left to right in Table~\ref{tab:dcaseTaxes_events}, the labels for D16T2 are all either identical (`clearing throat', `door knock', `door slam', `drawer', `phone ringing',`speech') or homographs and synonyms (`cough'=`coughing', `laughter'=`human laughter', `keyboard clicks'=`keyboard', `keys clinging'=`keys', `turning page'=`page turning') of Events in D13T2 and T3 so we do not reuse them. This process gives us 
$Ev_0=\{door knock, door slam, speech, laughter, keyboard, $
$impact, keys, phone, ringing, turning, page, cough, printer, $
$ alert, beep, short, throat, clear, mouse, click, drawer, switch\}$. 

Using the same process on D16T3 we produce \{Rustling, snapping, cupboard, cutlery, dishes, impact\} and some compound labels, \{glass jingling, people walking, washing dishes, and water tap running\}. The framework dismantles the compound labels into \{glass (obj), jingling (act), people (obj), walking (act), washing (act), we omit `dishes' as a duplicate, tap (obj), and running {act}\}. \\
$Ev_1=Ev_0+\{Rustling, snapping, cupboard, cutlery, dishes, $
$ impact, glass, jingling, people, walking, washing, tap, $
$ running\}$.


We continue in this fashion for all challenge label sets until we are left with:
$T_{events}=\{door, knock, slam, speech, laughter, keyboard, impact, keys, $
$ phone, ringing, turning, page, cough, printer, alert, beep, $
$ short, throat, clear, mouse, click, drawer, switch, rustling, $ 
$ snapping, cupboard, cutlery, dishes, impact, glass, jingling, $
$ people, walking, washing, tap, running, brakes, car, squeaking, $
$ children, large, vehicle, bird, singing, passing, shouting, wind, $ 
$ blowing, train, horn, air, truck, reversing, siren, fire, police,$
$ ambulance, engine, screaming, bike, skateboard, motorbike, $
$ dog, cat, frying, blender, vacuum, cleaner, shaver, toothbrush\}$. 
    
\begin{table}[h]
\centering \caption{Scene label set clusters from DCASE challenges (2013,2016-2018)}
\begin{tabular}{lll}
DCASE 13'(T1) & DCASE 16'\&17' (T1) &  DCASE 18' (T1) \\
\hline \hline
busy street 	& bus				& airport 					\\
quiet street 	& cafe / Restaurant & shopping mall		\\
park 			& car 				& metro station 			\\
open-air market & city center		& pedestrian street 		\\
bus 			& forest path		& public square				\\		
subway-train 	& grocery store		& street medium traffic		\\
restaurant 		& home				& tram (riding) 			\\
shop/supermarket & lakeside beach	& bus (riding) 				\\
office 			& library 			& metro (riding) 			\\
subway station 	& metro station		& urban Park 				\\
				& office 			&							\\
				& residential area 	&							\\
				& train 			&							\\
				& tram 				&							\\
				& urban park 		&							\\
\hline
\end{tabular}
\label{tab:dcaseTaxes_scenes}
\end{table}
In Table~\ref{tab:dcaseTaxes_scenes} we have listed label sets for each scene classification challenges from all DCASE workshops since 2013. Task 1 for DCASE 16 and DCASE 17 are identical therefore only listed once. The final set of labels produced with our framework which encompasses all previous labels is: \\
$T_{scenes}=\{bus, restaurant, shop, metro, airport, street, $ 
$supermarket, quiet, busy, park, office, station, car, city, $ 
$center, forest, pavement, library, train, tram, mall, public $
$space, riding\}$.
\\
\textbf{Finalising $T$}: we join $T_{events}$ and $T_{scenes}$ and remove duplicates between the two to form $T$. We further subselect the labels into the event, $ev$, environment, $en$, and context, $c$ subsets. $c$ is the smallest based on DCASE: $c=\{office, meeting, shopping\}$. 

\section{Summary}
\vspace{-0.5em}
In summary, with this extensible framework we hope to start an evolving taxonomy of classification labels for open set sound scene analysis. With this design, we enable machines to cross-classify as humans do; with consistent multiple taxonomies and using the cluster graph structure, this product enables correlation between dependent sound attributes in a scene, i.e. learning to discriminate the same event in different contexts.
If we ensure that future dataset labeling strategies build upon those which already exist such as expanding this taxonomy, then we can amalgamate datasets for future research. We aim to use our framework to align other datasets such as AudioSet with the proposed taxonomy. Although our approach is a small overhead when annotating new datasets, the long term benefits of data augmentation outweigh the cost, and our framework is much simpler than using other data migration methods (e.g. \cite{choe2015collective}) used in reusing datasets. 

Alongside sharing the initialised graph taxonomy, we have provided a central online point for links to future datasets which conform to the extensible approach so all researchers can link to the relevant parts/collections of datasets they wish to use for their own analyses\footnote{Email \texttt{casa.opentaxonomy@qmul.ac.uk} to add your dataset link}. A further benefit of this work is that it enables both bottom-up and top-down strategies for forming new label sets and sharing/combining them with other researchers. We hope that future DCASE challenge organisers will adopt this approach for managing labels in new datasets as this unification of sound labels will ease the annotation task.

%
%


\bibliographystyle{IEEEtran}
\bibliography{refs}

\begin{thebibliography}{10}
\providecommand{\url}[1]{#1}
\def\UrlFont{\rmfamily}
\providecommand{\newblock}{\relax}
\providecommand{\bibinfo}[2]{#2}
\providecommand\BIBentrySTDinterwordspacing{\spaceskip=0pt\relax}
\providecommand\BIBentryALTinterwordstretchfactor{4}
\providecommand\BIBentryALTinterwordspacing{\spaceskip=\fontdimen2\font plus
\BIBentryALTinterwordstretchfactor\fontdimen3\font minus
  \fontdimen4\font\relax}
\providecommand\BIBforeignlanguage[2]{{%
\expandafter\ifx\csname l@#1\endcsname\relax
\typeout{** WARNING: IEEEtran.bst: No hyphenation pattern has been}%
\typeout{** loaded for the language `#1'. Using the pattern for}%
\typeout{** the default language instead.}%
\else
\language=\csname l@#1\endcsname
\fi
#2}}

\bibitem{aly2005survey}
M.~Aly, ``Survey on multiclass classification methods,'' \emph{Neural
  Networks}, vol.~19, pp. 1--9, 2005.

\bibitem{salamon2014dataset}
J.~Salamon, C.~Jacoby, and J.~P. Bello, ``A dataset and taxonomy for urban
  sound research,'' in \emph{22nd ACM international Conference on Multimedia},
  2014, pp. 1041--1044.

\bibitem{gemmeke2017audio}
J.~F. Gemmeke, D.~P. Ellis, D.~Freedman, A.~Jansen, W.~Lawrence, R.~C. Moore,
  M.~Plakal, and M.~Ritter, ``Audio set: An ontology and human-labeled dataset
  for audio events,'' in \emph{IEEE International Conference on Acoustics,
  Speech and Signal Processing (ICASSP)}, 2017, pp. 776--780.

\bibitem{giannoulis2013detection}
D.~Giannoulis, E.~Benetos, D.~Stowell, M.~Rossignol, M.~Lagrange, and M.~D.
  Plumbley, ``{Detection and classification of acoustic scenes and events: An
  IEEE AASP challenge},'' in \emph{IEEE Workshop on Applications of Signal
  Processing to Audio and Acoustics (WASPAA)}, 2013, pp. 1--4.

\bibitem{battaglino2016open}
D.~Battaglino, L.~Lepauloux, and N.~Evans, ``The open-set problem in acoustic
  scene classification,'' in \emph{IEEE International Workshop on Acoustic
  Signal Enhancement (IWAENC)}, 2016, pp. 1--5.

\bibitem{macqueen1967some}
J.~MacQueen \emph{et~al.}, ``Some methods for classification and analysis of
  multivariate observations,'' in \emph{Proceedings of the fifth Berkeley
  symposium on mathematical statistics and probability}, vol.~1, no.~14, 1967,
  pp. 281--297.

\bibitem{aytar2016soundnet}
Y.~Aytar, C.~Vondrick, and A.~Torralba, ``Soundnet: Learning sound
  representations from unlabeled video,'' in \emph{Advances in Neural
  Information Processing Systems}, 2016, pp. 892--900.

\bibitem{gaver1993world}
W.~W. Gaver, ``What in the world do we hear?: An ecological approach to
  auditory event perception,'' \emph{Ecological Psychology}, vol.~5, no.~1, pp.
  1--29, 1993.

\bibitem{houix2012lexical}
O.~Houix, G.~Lemaitre, N.~Misdariis, P.~Susini, and I.~Urdapilleta, ``A lexical
  analysis of environmental sound categories.'' \emph{Journal of Experimental
  Psychology: Applied}, vol.~18, no.~1, p.~52, 2012.

\bibitem{guyot1997chapitre}
F.~Guyot, M.~Castellengo, and B.~Fabre, ``A study of the categorization of a
  household noise corpus,'' in \emph{Categorization and cognition: from
  perception to speech}, 1997, pp. 41--58.

\bibitem{vanderveer1980ecological}
N.~J. Vanderveer, \emph{Ecological acoustics: Human perception of environmental
  sounds}.\hskip 1em plus 0.5em minus 0.4em\relax Unpublished dissertation,
  1980.

\bibitem{schubert1975role}
E.~D. Schubert, \emph{The role of auditory perception in language
  processing}.\hskip 1em plus 0.5em minus 0.4em\relax York, 1975.

\bibitem{heittola2013context}
T.~Heittola, A.~Mesaros, A.~Eronen, and T.~Virtanen, ``Context-dependent sound
  event detection,'' \emph{EURASIP Journal on Audio, Speech, and Music
  Processing}, 2013.

\bibitem{pijanowski2011soundscape}
B.~C. Pijanowski, A.~Farina, S.~H. Gage, S.~L. Dumyahn, and B.~L. Krause,
  ``What is soundscape ecology? an introduction and overview of an emerging new
  science,'' \emph{Landscape Ecology}, vol.~26, no.~9, pp. 1213--1232, 2011.

\bibitem{amphoux1997paysage}
P.~Amphoux, ``Paysage sonore urbain,'' \emph{Le paysage et ses grilles,
  Françoise Chenet}, pp. 109--122, 1996.

\bibitem{simpson1989oxford}
J.~Simpson, E.~S. Weiner, \emph{et~al.}, ``{Oxford English dictionary
  online},'' \emph{Oxford: Clarendon Press. Retrieved March}, vol.~6, p. 2008,
  1989.

\bibitem{wordnet}
\BIBentryALTinterwordspacing
{Princeton University}, ``{About WordNet},'' 2010. [Online]. Available:
  \url{http://wordnetweb.princeton.edu/perl/webwn}
\BIBentrySTDinterwordspacing

\bibitem{west2001introduction}
D.~B. West \emph{et~al.}, \emph{Introduction to graph theory}.\hskip 1em plus
  0.5em minus 0.4em\relax Prentice Hall Upper Saddle River, 2001, vol.~2.

\bibitem{sridhar2010discovering}
M.~Sridhar, A.~G. Cohn, and D.~C. Hogg, ``Discovering an event taxonomy from
  video using qualitative spatio-temporal graphs,'' in \emph{19th European
  Conference on Artificial Intelligence (ECAI)}, vol. 215, 2010, pp.
  1103--1104.

\bibitem{bai2017automatic}
Y.~Bai, K.~Yang, W.-Y. Ma, and T.~Zhao, ``Automatic dataset augmentation,''
  \emph{arXiv preprint arXiv:1708.08201}, 2017.

\bibitem{enderton1977elements}
H.~B. Enderton, \emph{Elements of set theory}.\hskip 1em plus 0.5em minus
  0.4em\relax Academic Press, 1977.

\bibitem{copyrightFree}
\BIBentryALTinterwordspacing
pexels.com, ``Copyright free images,'' 2018. [Online]. Available:
  \url{https://www.pexels.com/search/conference/}
\BIBentrySTDinterwordspacing

\bibitem{mesaros2018detection}
A.~Mesaros, T.~Heittola, E.~Benetos, P.~Foster, M.~Lagrange, T.~Virtanen, and
  M.~D. Plumbley, ``{Detection and classification of acoustic scenes and
  events: Outcome of the {DCASE} 2016 challenge},'' \emph{IEEE/ACM Transactions
  on Audio, Speech and Language Processing (TASLP)}, vol.~26, no.~2, pp.
  379--393, 2018.

\bibitem{mesaros2017dcase}
A.~Mesaros, T.~Heittola, A.~Diment, B.~Elizalde, A.~Shah, E.~Vincent, B.~Raj,
  and T.~Virtanen, ``{DCASE} 2017 challenge setup: Tasks, datasets and baseline
  system,'' in \emph{Workshop on Detection and Classification of Acoustic
  Scenes and Events {(DCASE)}}, 2017.

\bibitem{dcase2018}
\BIBentryALTinterwordspacing
``{DCASE} workshop and challenge,'' 2018. [Online]. Available:
  \url{http://dcase.community/workshop2018/index}
\BIBentrySTDinterwordspacing

\bibitem{choe2015collective}
S.~H. Choe and Y.~M. Ko, ``Collective archiving of soundscapes in
  socio-cultural context,'' \emph{iConference Proceedings}, 2015.

\end{thebibliography}

\end{sloppy}
\end{document}